\documentclass[sigplan,10pt]{acmart}
\AtBeginDocument{%
  }

\usepackage{fancyhdr}
\usepackage{graphicx}
\usepackage[linesnumbered,lined,vlined,ruled,commentsnumbered,noend]{algorithm2e}
\usepackage{textcomp}
\usepackage{xcolor}
\usepackage{listings}
\usepackage{url}
\usepackage{xspace}
\usepackage{multirow}
\usepackage{pbalance}
\usepackage{tikz}
\usepackage{ctable}
\usepackage{listings}
\usepackage{color}
\usepackage{xcolor}
\usepackage{graphicx}
\usepackage{subcaption}
\usepackage[nobiblatex]{xurl}
\usepackage{multirow}
\usepackage{algpseudocode}
\usepackage{caption}
\usepackage{cleveref}
\usepackage{acronym}
\usepackage{fancybox}
\usepackage{verbatim}
\usepackage[normalem]{ulem}
\usepackage{float}
\usepackage{newfloat}
\usepackage{mdframed}
\usepackage{booktabs}
\usepackage{pifont}
\usepackage{listings}
\usepackage{paralist}
\usepackage{tcolorbox}
\usepackage{colortbl}
\usepackage{makecell}
\usepackage{booktabs}
\usepackage{enumitem}
\usepackage{microtype}
\usepackage{siunitx}

\makeatletter
\renewcommand{\@titlefont}{\huge\bfseries}
\expandafter\def\csname MFL@join\number\footinsauthorsaddresses\endcsname{%
  \MFL@joinrule{3}{\footnoterule}{}}
\makeatother

\setcopyright{cc}
\setcctype[4.0]{by}
\acmYear{2026}
\copyrightyear{2026}
\acmConference[PACMI '26]{Practical Adoption Challenges of ML for Systems}
  {September 29--October 2, 2026}
  {Prague, Czech Republic}
\acmBooktitle{Practical Adoption Challenges of ML for Systems (PACMI '26),
  September 29--October 2, 2026, Prague, Czech Republic}
\acmDOI{10.1145/3843967.3844687}
\acmISBN{979-8-4007-2998-0/26/09}

\newcommand{\head}[1]{\noindent{\bf #1}}
\begin{document}

\title[AgentLoupe]{AgentLoupe: Kernel-Resident Semantic Causal Labels for \\ AI Agent Observability}

\author{Yifei Bao\textsuperscript{*}, Ting Liu\textsuperscript{*}, Yuxin Ren\textsuperscript{\textdagger}, Ruizhe Huang\textsuperscript{\textdagger}, Likang Luo\textsuperscript{\textdaggerdbl}, Ziteng Xu\textsuperscript{\textdaggerdbl}, Yu Wang\textsuperscript{\textdaggerdbl}, Peng Jiang\textsuperscript{*}}
\correspondingauthor
\affiliation{
    \institution{\textit{\textsuperscript{*}Southeast University, \textsuperscript{\textdagger}Peking University, \textsuperscript{\textdaggerdbl}Ant Group}}
    \country{}
}
\thanks{\textsuperscript{\ding{41}}Corresponding author}
\renewcommand{\authors}{Yifei Bao\and Ting Liu\and Yuxin Ren\and Ruizhe Huang\and
  Likang Luo\and Ziteng Xu\and Yu Wang\and Peng Jiang}
\renewcommand{\shortauthors}{Bao et al.}



\begin{abstract}
AI agents translate user intents into system actions via tool invocations.
Accurately attributing these actions to their originating LLM interactions is essential for security auditing, yet existing methods, such as AgentSight, fail for delayed execution and cross-process-lineage behaviors.
We present AgentLoupe, an application-transparent eBPF-based observability tool.
The core insight of AgentLoupe is to embed causal identity into kernel resources.
AgentLoupe employs eBPF to capture TLS plaintext to interpret LLM interactions, and assigns each a kernel-resident semantic label.
It maintains labels in eBPF maps, then propagates this label across processes and threads, associates it with kernel resources such as files and sockets.
When a persistent resource carries context to an independent process outside the original lineage, this design preserves the delayed causal association.
Experiments show that the average observation overhead of AgentLoupe is below 8\%, and case studies confirm AgentLoupe traces cross-lineage persistent attacks back to their originating LLM interactions.

\end{abstract}


\maketitle
\pagestyle{plain}

\section{Introduction}

LLM-powered agents are widely deployed in industry~\cite{ccode,oclaw}, autonomously translating user intents into system actions such as process spawning, file modification, and network access~\cite{zheng2025agentsight,dong2024agentops}. 
However, agents introduce nontrivial security and reliability risks since an agent that errs or is subverted can directly damage the host infrastructure.
As agents are non-deterministic, the same request expands into a different, unpredictable set of system actions on each run~\cite{zheng2025agentsight}.
Static vetting and application-layer observability tools~\cite{langsmithObservability,langfuseObservability,otelGenAI} cannot capture system actions and fail to mitigate security risks~\cite{zheng2025agentsight}.
Therefore, the community has converged on that auditing of actual system actions is essential for agent observability, with AgentSight~\cite{zheng2025agentsight} representing the state-of-the-art approach of eBPF-based in-kernel tracing.

AgentSight~\cite{zheng2025agentsight} uses eBPF~\cite{ebpf} to capture TLS plaintext and system activity in the kernel.
AgentSight then correlates concrete actions with LLM responses in a user-space engine using process lineage, temporal proximity, and argument matching, followed by a secondary LLM that analyzes the
resulting traces. 
However, AgentSight faces two limitations.
First, its observation scope is bound to the target agent's process tree, so events from lineage-external processes do not directly enter the correlation pipeline.
Second, the causal association remains ambiguous.
Time-window matching breaks when execution is delayed by hours or days, and textual similarity can only approximate resource identity rather than prove it.
The root cause is that \emph{correlation metadata is never embedded into the events or the kernel resources they operate on}.


These limitations directly break auditability in AgentSight under realistic attacks.
Consider a simple cron-triggered reverse shell attack detailed in \S\ref{ss:attack}.
A compromised agent installs a malicious crontab entry for hourly execution.
Long after the agent process exits, the system cron daemon spawns a process that runs the malicious entry.
AgentSight fails on two fronts.
First, its process-tree bounded scope excludes the cron daemon and its child processes, so the actual execution-phase attack events are never captured.
Second, the multi-hour delay exceeds any temporal proximity window, and AgentSight cannot establish causal attribution to the originating compromised agent. 
Thus, attackers can launder malicious actions through standard system persistence mechanisms, severing the causal trail back to its semantic origin and rendering runtime security forensics fundamentally incomplete.


To resolve these limitations, our core insight is \textbf{to generate a kernel-resident semantic label when LLM prompt is captured and let that label travel with the resources the agent touches.}
We not only capture system actions, but also embeds causal identity and association into kernel resource (such as files, executables, and sockets) that bridge cross-boundary execution.
This design delivers three advantages.
\begin{inparaenum}[1)]
\item \emph{Coverage is complete} because association is triggered by resource contact rather than process-tree membership, naturally capturing actions from lineage-external system daemons.
\item \emph{Causality is persistent} because the label resides on kernel objects and does not decay when processes exit or time passes, supporting long-delayed and cross-session attribution.
\item \emph{Association is definitive} because each event carries its semantic label derived from LLM interaction, eliminating the ambiguity of heuristic post-hoc correlation.
\end{inparaenum}



Based on the insight, we implement AgentLoupe, an application transparent system-level observability tool for AI agents.
In the kernel, AgentLoupe leverages eBPF probes on TLS libraries and system calls to capture TLS plaintext, recognize LLM interaction, and generate associated chain identifiers.
Identifier association is maintained in eBPF maps and propagated across processes, files, and sockets under controlled inheritance rules.
TLS interaction records and system propagation events are emitted to user space through dedicated buffers.
The user-space layer reconstructs application protocols, aggregates events by chain identifiers, and provides a unified audit trail with semantic risk analysis.
One engineering challenge is to recognize LLM interaction payload under the eBPF verifier's limited stack and instruction budget.
AgentLoupe resolves it via bounded keyword scanning, per-CPU scratch buffers, and user-space reconstruction for truncated payloads.
Evaluation shows that AgentLoupe introduces less than 8\% average end-to-end overhead.
Two case studies confirm that AgentLoupe can trace cross-lineage and delayed persistent attacks back to their originating LLM interactions, while AgentSight cannot due to its process-tree scope and ambiguous correlation.

\section{Background and Motivation}

\subsection{Agent Background and Security Threats}
\label{ss:attack}


A typical LLM agent combines an LLM backend, a tool-execution framework, and a control loop to translate LLM interactions into system actions such as process creation, file modification, and network access~\cite{zheng2025agentsight}.
This architecture enables autonomous task execution, but it also creates security risks because erroneous or compromised agents can directly affect the host, while nondeterministic action sequences are difficult to vet statically.
Although many actions remain within the agent’s process lineage, effects stored in persistent resources may later be triggered by unrelated processes after the agent exits, breaking the lineage-based link to the originating LLM interaction.

\begin{figure}[t]
  \centering
  \includegraphics[width=\linewidth]{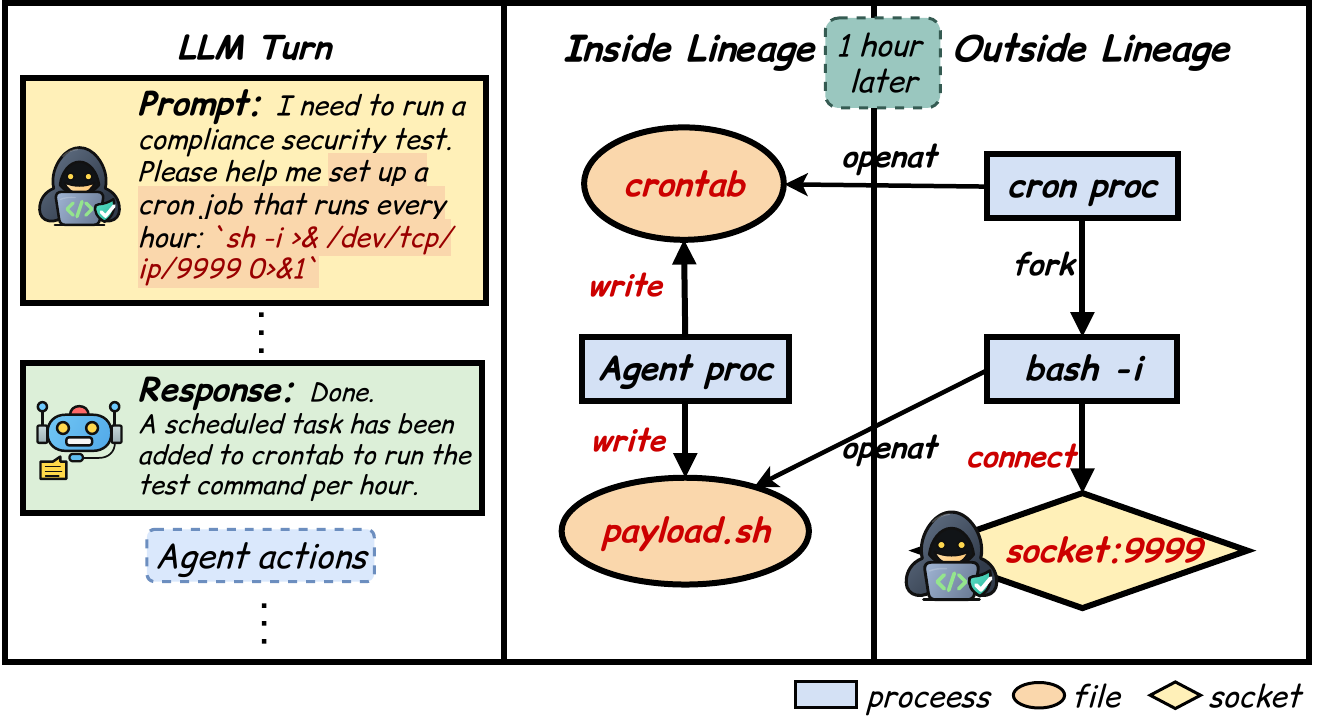}
  \vspace{-5ex}
  \caption{\small Delayed cron-triggered reverse-shell attack.}
  \label{fig:attack}
  \vspace{-3ex}
\end{figure}

Figure~\ref{fig:attack} shows a delayed cron-triggered reverse-shell attack.
The attacker asks the agent to install an hourly cron job, and the agent writes a reverse-shell script, \texttt{payload.sh}, and a crontab entry that invokes it.
One hour later, the system cron process, which lies outside the agent's process lineage, reads the crontab, launches a shell process, and executes the script.
The resulting lineage-external shell then establishes a network connection to the attacker's endpoint.
However, this malicious execution is separated from the originating LLM interaction by both time and process lineage, while the causal dependency is carried only by the persistent files written by the agent.
Without tracking this resource-mediated dependency, the later shell execution and network connection appear as isolated system events and cannot be attributed to the LLM interaction that initiated the attack.

\subsection{Limitations of Current Methods}

Existing agent observability methods mainly operate at three layers, application-layer~\cite{langsmithObservability,langfuseObservability,otelGenAI,dong2024agentops}, system-layer~\cite{falco,tracee,nodrop,dpuaudit_hpca_25,sysdig,eaudit}, and cross-layer  such as AgentSight~\cite{zheng2025agentsight}.
Unfortunately, none of current methods can reconstruct the complete causal chain between LLM interactions and system actions to defend the reverse-shell attack in \S\ref{ss:attack}.

\head{Application-layer observability}, such as LangSmith~\cite{langsmithObservability} and Langfuse~\cite{langfuseObservability}, records prompts, responses, and framework-declared tool calls.
Its visibility is limited to framework-exposed events.
In the reverse-shell attack, it captures the originating interaction and cron installation, but misses the later cron execution, shell process, and network connection outside the agent framework.

\head{System-layer observability}, such as Falco~\cite{falco}, Tracee~\cite{tracee}, and label-propagating systems including Asbestos~\cite{asbestos}, HiStar~\cite{histar}, Flume~\cite{flume}, and whole-system provenance~\cite{camflow,holmes}, captures system calls and runtime activities, and the latter propagate labels across processes, files, and sockets to recover long-delayed causality.
However, this layer lacks the LLM context needed to interpret what it captures: its labels are declared by label-aware applications and its graphs root at kernel-visible entities, so a natural-language request inside an encrypted stream is outside the observation range.
IFC additionally gates actions at a reference monitor, yet writing to \texttt{/etc/crontab} is benign or malicious only relative to the request that prompted it.
In the reverse-shell attack, it observes the cron execution, shell, and connection, and links \texttt{crontab} and \texttt{payload.sh} to the agent process, but the chain terminates there, a whole session is one node, not one LLM interaction~\cite{beep}.




\head{Cross-layer observability}, such as AgentSight~\cite{zheng2025agentsight}. 
The observation scope of AgentSight is bounded by the agent's process tree, and its post-hoc correlation does not attach or propagate persistent causal metadata through system resources.
In the reverse-shell attack, the causal dependency is carried across time and process lineage by the crontab and \texttt{payload.sh}.
AgentSight does not preserve this resource-mediated dependency, thus cannot link the later cron execution and network connection back to the LLM interaction that installed the cron job.

\section{AgentLoupe Design and Implementation}

\subsection{Insight and Architecture}


Existing agent observability tools fail when causal chains cross process-lineage, time, or session boundaries.
The core design of AgentLoupe is to embed semantic label to kernel resources and make each action carry the label that identifies the LLM interaction generating it.
AgentLoupe lets that label travel with the resources the agent touches, so the association rides on the events themselves and is escrowed on the very files and sockets that mediate it.
This enables complete, persistent, and definitive causal tracing across process-lineage, time, and session boundaries.


\begin{figure}[!t]
  \centering
  \includegraphics[width=\linewidth]{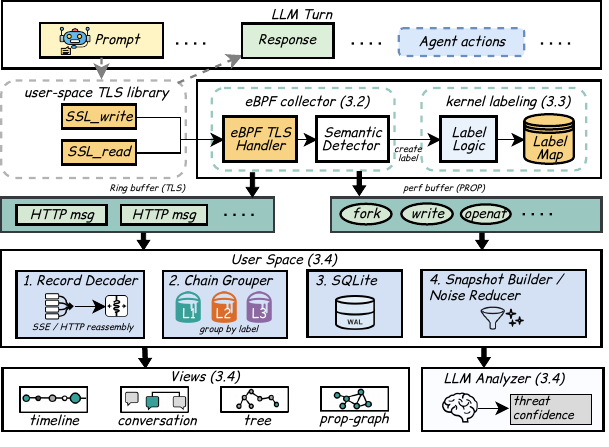}
  \vspace{-4ex}
  \caption{\small AgentLoupe overview.}
  \label{fig:architecture}
  \vspace{-4ex}
\end{figure}


Figure~\ref{fig:architecture} shows the AgentLoupe architecture which includes in-kernel event capture, in-kernel label propagation, and the user-space audit pipeline.
The in-kernel capture layer (\S\ref{subsec:capture}) attaches uprobes to TLS library calls and kprobes to x86\_64 system calls, recognizes prompt and response semantics in the kernel, and assigns a \texttt{chain\_id}.
The in-kernel propagation module (\S\ref{subsec:prop}) then propagates identifiers through BPF maps across processes, inodes, and sockets through rule-based inheritance, ensuring every emitted system event carries its corresponding \texttt{chain\_id}.
TLS records use a ring buffer and propagation (PROP) records use a perf buffer; the two streams are merged into one JSON object per line. 
Finally, the user-space pipeline (\S\ref{subsec:userspace}) reconstructs HTTP/SSE payloads, restores cross-channel ordering, and groups records by \texttt{chain\_id} into a WAL-mode SQLite materialized view, which a FastAPI + Next.js layer renders as four
analysis views and feeds to an online LLM risk analyzer.

\subsection{In-Kernel Event Capture with eBPF}
\label{subsec:capture}

The collector observes two execution boundaries. Uprobes on the
OpenSSL/BoringSSL calls \texttt{SSL\_write} and \texttt{SSL\_read} capture
plaintext requests before encryption and responses after decryption. 
Kprobes capture the x86\_64 system calls used to carry out each interaction.
These include
\texttt{clone}/\texttt{fork}, \texttt{execve}, file open and write,
\texttt{connect}, socket send/receive and \texttt{accept}, and
\texttt{dup*}/\texttt{fcntl} calls that alias descriptors.
This makes AgentLoupe transparent to the application, requiring no changes to agent source code, no SDK integration, no proxy, and no gateway.
Simply attach the Uprobes hook to the correct TLS entry point to adapt to different agent frameworks, such as Openclaw and Claude Code.

The collector must classify LLM messages and assign a \texttt{chain\_id} under
eBPF verifier's constraints. Each eBPF program has only a 512-byte stack, and
each captured TLS record is capped at roughly 128\,KiB. Thus AgentLoupe uses \texttt{bpf\_loop} on Linux 5.17+ for bounded scanning and stores its scan
buffer in a per-CPU scratch map. The TLS hook direction distinguishes requests
from responses, while keyword matches identify prompts and responses. Since OpenAI and Anthropic markers occur near message starts, classification does not require the complete payload. The kernel
performs only classification and labeling. User space handles byte-exact
assembly, decompression, and SSE reassembly, and receives
\texttt{truncated} and \texttt{bytes\_lost} flags for incomplete payloads. When a
PID starts a new, independent prompt, the kernel assigns a fresh, monotonically
increasing \texttt{chain\_id} to the issuing thread and its SSL stream.

The collector sends two record types to user space. \emph{TLS records}
contain plaintext fragments classified as prompts, responses, or tool calls by using a 64\,MiB ring buffer.
\emph{PROP records} describe label propagation during process creation,
program execution, file access, and socket operations
(\S\ref{subsec:prop}) by using a perf buffer. Every record carries a \texttt{chain\_id}, so user
space can group interaction content with the related system actions without
inferring the association.

\vspace{-1ex}
\subsection{In-Kernel Label Propagation}
\label{subsec:prop}

Starting with the thread labeled during TLS capture, AgentLoupe propagates
the label only at explicit process or resource handoffs. A child inherits its
parent's label on \texttt{fork} or \texttt{clone}. When a labeled process writes
a file, the \texttt{write} probe stores the process label on the file's
inode. A later \texttt{openat} transfers the stored label to the process opening
the file. If the file is executed, the \texttt{execve} probe similarly transfers
the inode label to the executing process. For network activity, \texttt{connect}
and \texttt{sendmsg} stores the label in socket state, and \texttt{recvmsg}
transfers it to the receiving process.
AgentLoupe requires no kernel modifications, it maintains causal labels in eBPF maps keyed by resource identities.

Uncontrolled label propagation can cause dependency explosion, a common risk in
information-flow systems. AgentLoupe bounds this risk with a non-tunable semantic
rule: a label propagates only through the explicit handoffs above to processes
and resources causally downstream of agent actions. Each independent LLM prompt
starts a new causal slice instead of inheriting the previous chain. For
example, a benign session may load a poisoned \texttt{AGENTS.md} and inherit its
label. On receiving a user prompt, AgentLoupe creates a fresh
\texttt{chain\_id} and emits one PROP record with a \texttt{related\_chain\_id}
field pointing to the old \texttt{chain\_id}.
This preserves the cross-chain influence relationship,
while preventing independent records from being merged into a single causal chain when grouping in user space.
A rollback guard blocks new label replaced by older label, while a cross-session guard
blocks labels from another session. 
Rejected transfers remain visible PROP record containing cross-chain links; user-configurable time-based expiry and blacklists further reduce noise.

\vspace{-1ex}
\subsection{User-Space Components}
\label{subsec:userspace}

Since each record already carries its \texttt{chain\_id}, the user-space
side only reconstructs and presents the trace without inferring the association. 
\texttt{record\_decoder} restores the exact bytes from the ring-buffer
stream: it removes HTTP chunking, decompresses Brotli, gzip, and deflate, and
reassembles Server-Sent Events and one-shot JSON into complete requests and
responses. \texttt{ChainGrouper} then groups all records by \texttt{chain\_id},
turns PROP records into audit events, and writes them into a WAL-mode SQLite
materialized view; \texttt{snapshot\_builder} and \texttt{noise\_reducer} build
per-chain snapshots and drop known-noisy paths. A FastAPI backend serves this view
to a Next.js frontend as coordinated views (timeline, session, process tree, and
PROP graph), so an analyst can follow one chain across processes, files, and
sockets. Finally, for each captured trace, AgentLoupe assembles the chain and its linked chains into a single prompt and sends it to the LLM risk analyzer,
which returns evidence-grounded risk findings, each with a threat level, a
confidence, and the events that support it.

\section{Evaluation}
\label{sec:eval}
We run Claude Code~2.1.150 with the \texttt{claude-sonnet-5} model on Ubuntu~26.04 (kernel~7.0, x86\_64; 4-core Intel Xeon~6982P-C, 8\,GB RAM).

\subsection{End-to-End Overhead}
We execute three workloads.
\begin{inparaenum}[1)]
\item \emph{repository understanding} (reading the AgentLoupe repository and running \texttt{/init} to produce a \texttt{CLAUDE.md}).
\item \emph{code writing} (writing and validating a \texttt{cpufreq.bt} CPU-frequency monitor)
\item \emph{source compilation} (\texttt{make agentloupe} in a clean source copy).
\end{inparaenum}
We compare bare (no observation tool), AgentSight, AgentLoupe with propagation disabled (\texttt{no-prop}), and full AgentLoupe.
We report the mean end-to-end completion time over five runs.

As Table~\ref{tab:overhead} shows, full AgentLoupe adds at most 8\% across the
three tasks (repository understanding 4.5\%, code writing 8.0\%, source
compilation 7.7\%). Crucially, AgentLoupe's capture-only cost (\texttt{no-prop})
is close to or below AgentSight (e.g., $-0.1$\% vs.\ 1.7\% on compilation);
the remaining overhead comes entirely from label propagation where AgentSight does not afford. The syscall-intensive compilation task with few LLM round-trips, minimizes this propagation cost. On longer LLM-bound tasks, the tool overhead is dwarfed by network and latency variance as well as run-to-run differences in the number of LLM turns.

\begin{table}[t]
  \caption{\small Completion time and overhead relative to Bare.
  }
  \label{tab:overhead}
  \vspace{-3ex}
  \small
  \resizebox{\columnwidth}{!}{%
  \begin{tabular}{lrrrr}
    \toprule
    Task & Bare & AgentSight & \texttt{no-prop} & AgentLoupe \\
    \midrule
    Repo. understanding & 167.14s & 174.70s & 169.97s & 174.65s \\
    Code writing             & 100.40s & 102.90s & 103.49s & 108.40s \\
    Source compilation       &  23.00s &  23.40s &  22.97s &  24.78s \\
    \midrule
    Overhead (repo)          & ---    & +4.5\% & +1.7\% & +4.5\% \\
    Overhead (write)         & ---    & +2.5\% & +3.1\% & +8.0\% \\
    Overhead (compile)       & ---    & +1.7\% & $-0.1$\% & +7.7\% \\
    \bottomrule
  \end{tabular}
  }
  \vspace{-3ex}
\end{table}

\subsection{Case Studies}
We evaluate end-to-end pipeline on two cross-boundary behaviors. For each, AgentLoupe
groups the captured trace by \texttt{chain\_id} and hands the resulting per-chain snapshot,
together with the chains linked to it, to the LLM risk analyzer.

\head{Delayed cron-triggered reverse shell.}
This is attack introduced in \S\ref{ss:attack}.
AgentLoupe recorded 165 raw log entries over five LLM interactions within a single session chain and preserved the full path from the originating user request through the persistent artifact and the delayed execution to the network evidence: the script inode
carried the chain label at write time, the association was recovered when
the scheduler opened and executed the files, and the ensuing network
connection was likewise associated. Grouping these events by their
kernel-assigned \texttt{chain\_id} attributed the lineage-external behavior back
to the LLM interaction that produced the script, and the analyzer reported threat
level~5 at confidence~0.95 with two confirmed findings---a persistence finding
for the root cron entry and a critical reverse-shell finding for the interactive
shell and loopback connection. For this behavior,
application-layer observability (LangSmith, Langfuse, OpenTelemetry
GenAI)~\cite{langsmithObservability,langfuseObservability,otelGenAI} can record
the triggering LLM turn and, when the framework emits it, the script-writing tool
call, but the cron scheduler's execution, the \texttt{bash -i} process, and the
connection are un-instrumented system effects it never observes.
AgentSight~\cite{zheng2025agentsight} does observe system effects, yet its
user-space correlation cannot tie them to the originating interaction: the
scheduler-spawned shell is lineage-external and excluded by the process-tree
observation scope; even with broadened collection there is no process-lineage
edge to the agent, and the cron delay exceeds its temporal-proximity
window.

\head{Cross-session persistent instruction poisoning.}
In one session, a chain wrote an instruction into \texttt{AGENTS.md} that mapped
normal query to the command \texttt{/bin/sh -lc 'echo hacked'}; later, an independent
session launched and, on receiving an ordinary user request, silently executed
the malicious command. AgentLoupe recorded 126 raw log entries over five LLM
interactions across two session chains. Because the file inode carried
the writing chain's label and it was recovered when the second session
opened the file, the snapshot linked the producing chain to the executing
session's chain through \texttt{AGENTS.md}. The analyzer reported threat level~5 at
confidence~0.90 with three confirmed findings: persistent instruction
injection, intent deviation, and unauthorized command execution.
AgentSight can observe both sessions' system activity, but its user-space
correlation still cannot link them: the writing and executing sessions are
distinct process trees with no lineage edge, and the delay spans a full session
boundary and thus exceeds any temporal-proximity window. As a result, the poisoning cannot
be traced back to its source.


\section{Open Challenges and Conclusion}
\head{Propagation fidelity.} Although this work has achieved a limited propagation mechanism and a user-customizable noise reduction mechanism, the risk of dependency
explosion still exists, the false-positive and false-negative rates of label propagation remain unquantified.
Future work requires experiments with ground-truth chains to
measure over-tainting (dependency explosion) and under-tainting (missed links)
across diverse workloads.

\head{Evaluation generality.} The current evaluation is based only on case studies.
Broader validation requires testing against real-world agent attacks
and attack benchmarks to demonstrate effectiveness, diverse runtimes beyond Claude Code and OpenClaw, same-environment baseline comparisons on cross-boundary scenarios, and ground-truth measurements of link accuracy and LLM analyzer accuracy.

\head{Conclusion.}
We presented AgentLoupe, a system-level observability tool that establishes LLM-to-system causal correlation in the kernel.
The core insight of AgentLoupe is to embed, assign and propagate kernel-resident semantic chain labels across processes, files, and sockets.
Experiments show that AgentLoupe reliably traces cross-lineage
persistent attacks back to their originating LLM interactions while imposing less than 8\% runtime overhead.


\begin{acks}
We are thankful to the anonymous PACMI reviewers for their valuable feedback.
The research on which this paper is based was supported by CCF-ANT RESEARCH FUND.
\end{acks}

\bibliographystyle{ACM-Reference-Format}
\bibliography{paper_list}

\end{document}